\documentclass{jps-cp}
\usepackage{txfonts} %Please comment out this line unless the txfonts package is availabe in your LaTeX system.
\usepackage{color}

\title{Derivation of Interacting Two-Qubit Dynamics from Spin-Boson Model}

\author{Hiroaki \textsc{Matsueda}$^{1,2}$, Yukiya \textsc{Ide}$^{1}$, and Sadamichi \textsc{Maekawa}$^{3,4}$}

\inst{$^{1}$Department of Applied Physics, Graduate School of Engineeering, Tohoku University, Sendai 980-8579, Japan\\
$^{2}$Center for Science and Innovation in Spintronics, Tohoku University, Sendai 980-8577, Japan \\
$^{3}$RIKEN Center for Emergent Matter Science, Wako, Saitama 351-0198, Japan\\
$^{4}$Kavli Institute for Theoretical Physics, University of Chinese Sciences, Beijing 100190, China}

\email{hiroaki.matsueda.c8@tohoku.ac.jp}

\recdate{July 31, 2022}

\abst{We derive damping equations of motion for interacting two-spin states from a spin-boson model in order to examine qubit dynamics in quantum computers. On the basis of the composite operator method, we develop the Caldeira-Leggett approach for open quantum systems so that the entanglement dynamics originated from the two-spin correlation can be taken. We demonstrate numerical results for time dependence on the two-spin dynamics. We find that the relaxation of the total spin is described by a quantum version of the Landau-Lifshitz-Gilbert equation for magnetic materials. We also find that a two-spin composite mode keeps oscillation even after the total spin has been fully relaxed. We thus conclude that the two-spin correlation due to the presence of the composite mode is stable against dissipation. We consider the mechanism of why the correlation is maintained.}

\kword{qubit, spin-boson model, relaxation, composite operator, equation of motion, Caldeira-Leggett approach}

\begin{document}
\maketitle

\section{Introduction}

Nowadays, exploring quantum technologies such as quantum computation, quantum cryptography, and quantum sensing is turning into a realistic goal for current engineering. Ten years have passed already after the D-wave machine, a kind of quantum annealers, was commercially provided, and the machine is getting used for various optimization problems. Recent flagship research projects associated with quantum technologies aim to construct large-scale fault-tolerant universal quantum computers in the middle of this century. For this purpose, development of noisy intermediate-scale quantum (NISQ) computers is an important milestone at the present stage. We believe that future sophisticated society will be highly supported from these technologies, and thus it is necessary to promote basic science behind the technologies. There are mainly two directions for the advanced research. One is direct treatment of fault tolerance algorithm and implementation as long-term research, and the other is deep examination of NISQ itself as short or medium-term plan. The latter is closely related to non-equilibrium physics in which the dynamics of our qubit system is highly disturbed by the environmental noise and interaction among qubits themselves when the qubits are massively integrated on the substrate. Since a qubit can be identified with a quantum spin, our target model is the so-called spin-boson model in which the interacting quantum spins couple with bosonic degrees of freedom. Thus, our interest is to understand operational stability of single spin and entanglement among multiple spins in this model. This is because quantum computation is realized by sequential change of qubit states with external perturbation as unitary gates.

Motivated by the abovementioned consideration, we theoretically examine the spin dynamics in the spin-boson model. Here, the entanglement control of qubits is a key for various quantum technologies, and thus we particularly focus on whether dynamical behavior of non-local correlation or entanglement is stably controlled against dissipation due to the presence of the environment. This type of works was recently done as a toy model for the D-wave machine~\cite{ref1}. However, this is a very special case in which we can successfully integrate out bosonic degrees of freedom in terms of the Suzuki-Trotter decomposition. We would like to get versatile techniques for more general cases.

For this purpose, we first derive damping equations of motion for the total spin from our spin-boson model in which two spins interact with each other. We developed old approaches such as Feynman's influential functional and Caldeira-Leggett model. We assume Ohmic spectral distribution of environmental degrees of freedom and Markovian approximation to get a closed equation of motion. We then find that the result is equivalent to a quantum version of the Landau-Lifshitz-Gilbert (LLG) equation for macroscopic spin precession in magnetic materials~\cite{ref2,ref3}. Here, the Gilbert constant is proportional to the coefficient of the distribution function. On the basis of the total-spin dynamics, we next focus on the internal dynamics of two spins in order to understand the stability of their nonlocal correlation. The abovementioned theoretical method is also applied to a composite operator associated with the correlation. We find that the two-spin correlation is maintained even after the total spin has been relaxed and the stability of two-spin correlation is different from that of the total-spin dynamics~\cite{ref3}. In the usual spectroscopy in quantum many-body systems, low-lying states are dominated by composite spins~\cite{ref4}. We expect that the stability of the dynamics strongly depends on the spatial size of correlated spin cloud. We demonstrate numerical results and mention why the stability of the nonlocal correlation appears.

\section{Model and Equations of Motion for Interacting Two-Spin Dynamics}

We consider interacting two qubits (quantum spins) coupled to the bosonic environment. We start with the following Hamiltonian
\begin{eqnarray}
H=\sum_{k,\alpha}\omega_{k}b_{k}^{\alpha\dagger}b_{k}^{\alpha}+\sum_{\alpha}B_{0}^{\alpha}S^{\alpha}+J\sum_{\alpha}S_{1}^{\alpha}S_{2}^{\alpha}+\sum_{k,\alpha}\left(\nu_{k}b_{k}^{\alpha}+\nu_{k}^{\ast}b_{k}^{\alpha\dagger}\right)S^{\alpha},
\end{eqnarray}
where we consider two $S=1/2$ spins, $S^{\alpha}=S_{1}^{\alpha}+S_{2}^{\alpha}$ ($\alpha=1,2,3$), $b_{k}^{\alpha\dagger}$ and $b_{k}^{\alpha}$ are boson operators with mode $k$ and index of angular momentum $\alpha$, $B_{0}^{\alpha}$ is associated with the energy difference of qubit states, $\epsilon$, and transverse field, $h$, $\vec{B}_{0}=(h,0,\epsilon)$, and $J$ is antiferromagnetic coupling between spins. Here we assume a special form of boson operators with index $\alpha$ so that we can find simple damping equations for spins. Furthermore, the coupling between qubits is assumed to be of Heisenberg type for simplicity, but this is an ideal situation. We should note that the coupling actually depends on the type of qubit design. For simplicity we introduce
\begin{align}
A^{\alpha}=B_{0}^{\alpha}+L^{\alpha}\; , \; L^{\alpha}=\sum_{k}\left(\nu_{k}b_{k}^{\alpha}+\nu_{k}^{\ast}b_{k}^{\alpha\dagger}\right),
\end{align}
and then the Hamiltonian is simply represented as $H=\sum_{k}\omega_{k}\vec{b}_{k}^{\dagger}\cdot\vec{b}_{k}+J\vec{S}_{1}\cdot\vec{S}_{2}+\vec{A}\cdot\vec{S}$.

The equations of motion for single spin operators, $\vec{S}_{1}$ and $\vec{S}_{2}$, are represented as
\begin{align}
\frac{\partial}{\partial t}\vec{S}_{1}=J\vec{m}-\vec{S}_{1}\times\vec{A} \; , \; \frac{\partial}{\partial t}\vec{S}_{2}=-J\vec{m}-\vec{S}_{2}\times\vec{A},
\end{align}
where the composite spin operator $\vec{m}$ is defined by
\begin{eqnarray}
m^{\alpha}=\epsilon_{\alpha\beta\gamma}S_{1}^{\gamma}S_{2}^{\beta} \; , \; \vec{m}=\vec{S}_{2}\times\vec{S}_{1},
\end{eqnarray}
and this operator characterizes entanglement between two spins. To understand the relation between $\vec{m}$ and entanglement, it is useful to remember the definition of the single spin $\vec{S}_{1}$. The states, $\left|\uparrow\right>$ and $\left|\downarrow\right>$, are the eigenstates of $S_{1}^{z}$ in the single-spin case. The transition between these states are represented by $S_{1}^{x}$ and $S_{1}^{y}$. Here, the eigenstates of $m^{z}$ are the product and entangled states ($\left|\uparrow\uparrow\right>$, $\left|\downarrow\downarrow\right>$, and $\left|\uparrow\downarrow\right>\pm i\left|\downarrow\uparrow\right>$). Then, $m^{x}$ and $m^{y}$ correspond to the transition from the product state ($\left|\uparrow\uparrow\right>$ or $\left|\downarrow\downarrow\right>$) to the singlet state, and these transition operators play a crucial role on generating or keeping entanglement. Thus, we particularly focus on the relaxation dynamics of $\vec{m}$. Note that the equation of motion for the total spin $\vec{S}$ does not include $\vec{m}$ because $\vec{S}$ commutes with the Heisenberg coupling:
\begin{eqnarray}
\frac{\partial}{\partial t}\vec{S}=-\vec{S}\times\vec{A}=-\frac{1}{2}\left(\vec{S}\times\vec{A}-\vec{A}\times\vec{S}\right). \label{total}
\end{eqnarray}
Thus, the alternative treatment of $\vec{m}$ is important for the examination of internal dynamics between spins. The equation of motion for $\vec{m}$ is given by
\begin{eqnarray}
\frac{\partial}{\partial t}\vec{m}=\frac{1}{2}J\left(\vec{S}_{2}-\vec{S}_{1}\right)-\vec{m}\times\vec{A}=\frac{1}{2}J\left(\vec{S}_{2}-\vec{S}_{1}\right)-\frac{1}{2}\left(\vec{m}\times\vec{A}-\vec{A}\times\vec{m}\right). \label{j1j2}
\end{eqnarray}
Note that we have taken symmetrized procedure in Eqs.~(\ref{total}) and (\ref{j1j2}) in order to avoid technical difficulty associated with noncommutativity of quantum operators.

These equations still contain bosonic operators through $\vec{A}$. Let us remove the bosonic degrees of freedom. For this purpose, the Heisenberg equation of motion for environmental boson is given by 
\begin{eqnarray}
i\frac{\partial}{\partial t}b_{k}^{\alpha}=\omega_{k}b_{k}^{\alpha}+\nu_{k}^{\ast}S^{\alpha},
\end{eqnarray}
and the formal solution can be obtained as
\begin{eqnarray}
b_{k}^{\alpha}(t)=e^{-i\omega_{k}t}b_{k}^{\alpha}(0)-i\nu_{k}^{\ast}\int_{0}^{t}dt^{\prime}e^{-i\omega_{k}\left(t-t^{\prime}\right)}S^{\alpha}\left(t^{\prime}\right).
\end{eqnarray}
We would like to obtain a closed form of equations of motion for $\vec{S}$ and $\vec{m}$, and for this purpose we substitute the bosonic solution into the equations. We assume the bosonic spectrum as
\begin{eqnarray}
J(\omega)=\sum_{k}\left|\nu_{k}\right|^{2}\delta\left(\omega-\omega_{k}\right)=\eta\omega,
\end{eqnarray}
where this assumption represents the Ohmic process and the coefficient $\eta$ plays a central role on the relaxation of spin dynamics.

By combining these equations with use of Markovian approximation ($\omega_{c}$ is the cut-off frequency for $J(\omega)$, and we take it as a large constant), the final form of total spin dynamics is given by
\begin{align}
\frac{\partial}{\partial t}\vec{S}(t)=&\vec{B}(t)\times\vec{S}(t)-\eta\frac{\sin\omega_{c}t}{t}\left(\vec{S}(t)\times\vec{S}(0)-\vec{S}(0)\times\vec{S}(t)\right) \nonumber \\
&-\alpha(t)\left(\vec{S}(t)\times\frac{\partial}{\partial t}\vec{S}(t)-\frac{\partial}{\partial t}\vec{S}(t)\times\vec{S}(t)\right), \label{LLGsingle}
\end{align}
where the coefficient $\alpha(t)$ is defined by
\begin{eqnarray}
\alpha(t)=\eta\int_{0}^{\omega_{c}t}d\tau\frac{\sin\tau}{\tau} \; , \; \lim_{\omega_{c}t\rightarrow\infty}\alpha(t)=\frac{\pi\eta}{2},
\end{eqnarray}
and $\vec{B}(t)$ is defined by
\begin{eqnarray}
B^{\alpha}(t)=B_{0}^{\alpha}+\sum_{k}\left(\nu_{k}e^{-i\omega_{k}t}b_{k}^{\alpha}(0)+\nu_{k}^{\ast}e^{i\omega_{k}t}b_{k}^{\alpha\dagger}(0)\right). \label{field}
\end{eqnarray}
The result is essentially a quantum version of the LLG equation with the damping coefficient $\alpha(t)$. In the right hand side of Eq.~(\ref{LLGsingle}), the second term represents correlation between the initial state and the state at time $t$. For large $\omega_{c}$ values, the second term becomes negligible with time. The last term in Eq.~(\ref{LLGsingle}) shows damping, and also produces quantum effects that do not contain in the classical LLG dynamics. The quantum effects are originated in non-commutativity between $\vec{S}$ and $\partial\vec{S}/\partial t$, and then the magnitude of the expectation value of $\vec{S}$ may not be a conserved quantity. We will briefly discuss this point later.

On the other hand, $\vec{m}$ shows the following dynamics
\begin{align}
\frac{\partial}{\partial t}\vec{m}(t)=& \vec{B}(t)\times\vec{m}(t)+\frac{1}{2}\left(J-2\eta\omega_{c}\right)\left(\vec{S}_{2}(t)-\vec{S}_{1}(t)\right) \nonumber \\
&-\eta\frac{\sin\omega_{c}t}{t}\left(\vec{m}(t)\times\vec{S}(0)-\vec{S}(0)\times\vec{m}(t)\right)-\alpha(t)\left(\vec{m}(t)\times\frac{\partial}{\partial t}\vec{S}(t)-\frac{\partial}{\partial t}\vec{S}(t)\times\vec{m}(t)\right). \label{mdot0}
\end{align}
By substituting $\vec{m}\times\vec{S}=\left(\vec{S}_{1}-\vec{S}_{2}\right)/2+i\vec{m}$ and $\vec{S}\times\vec{m}=\left(\vec{S}_{2}-\vec{S}_{1}\right)/2+i\vec{m}$ into Eq.~(\ref{mdot0}), we obtain
\begin{align}
\frac{\partial}{\partial t}\vec{m}(t)=&\vec{B}(t)\times\vec{m}(t)-\frac{1}{2}\left(J-2\eta\omega_{c}\right)\left(\vec{m}(t)\times\vec{S}(t)-\vec{S}(t)\times\vec{m}(t)\right) \nonumber \\
&-\eta\frac{\sin\omega_{c}t}{t}\left(\vec{m}(t)\times\vec{S}(0)-\vec{S}(0)\times\vec{m}(t)\right)-\alpha(t)\left(\vec{m}(t)\times\frac{\partial}{\partial t}\vec{S}(t)-\frac{\partial}{\partial t}\vec{S}(t)\times\vec{m}(t)\right). \label{mdot}
\end{align}
A striking feature of this equation is that the relaxation term (the last term in the right hand side) turns off when $\vec{S}$ is in a stationary condition $\partial\vec{S}/\partial t=0$. In this case, the operator $\vec{S}$ in the second term behaves as a static field. The third term represents quantum correlation between initial state and the state at time $t$, but this term is negligible as we have already pointed out. Furthermore, decoupling of the nature of $\vec{S}$ and $\vec{m}$ is facilitated by taking $J=2\eta\omega_{c}$. Therefore, the dynamics of $\vec{m}$ is very stable against the relaxation of $\vec{S}$. More precisely, the stationary condition must be represented by $\left<\psi\right|\partial\vec{S}/\partial t\left|\psi\right>=\partial\langle\vec{S}\rangle/\partial t=0$ with the initial quantum state $\left|\psi\right>$. Thus, the abovementioed statement would be too strong. We expect slow damping of the composite spin $\vec{m}$, and the damping behavior of $\vec{m}$ would be represented by sophisticated treatment of higher-order equation of motion. We thus think that the energy relaxation time $T_{1}$ is determined by the time scale of relaxation of the total spin $\langle\vec{S}\rangle$, and the decoherence time $T_{2}$ between two spins is determined by the time scale of relaxation of the composite spin $\langle\vec{m}\rangle$. In the next section, we analyze these coupled equations of motion to examine the feature of the dynamics of $\langle\vec{S}\rangle$ and $\langle\vec{m}\rangle$.

In the previos works associated with inertial spin dynamics in ferromagnets, higher-order terms of the LLG equation have been considered~\cite{ref5,ref6}. It is an interesting future work to examine their relationship with the present result.

\section{Numerical Results}

Before going into numerical details, we briefly examine the qubit system without bosons in order to find a guideline for determining the magnitudes of $J$ and $\epsilon$ ($h=0$). In this case, the Hamiltonian without bosons, $H_{0}$, is transformed into
\begin{eqnarray}
H_{0}=J\vec{S}_{1}\cdot\vec{S}_{2}+\epsilon S^{z}=\frac{J}{2}S(S+1)-\frac{3J}{4}+\epsilon S^{z},
\end{eqnarray}
where $\vec{S}\cdot\vec{S}=S(S+1)=2\vec{S}_{1}\cdot\vec{S}_{2}+3/2$ and thus $H_{0}$ can be represented by using the total spin $S$. The singlet is characterized by $S=0$ and $S^{z}=0$, while the triplet is characterized by $S=1$ and $S^{z}=-1,0,+1$. Here, we compare the energy of singlet $E(S=0,S^{z}=0)=-3J/4$ with the energy of one of triplets $E(S=1,S^{z}=-1)=J/4-\epsilon$. Then, $E(S=1,S^{z}=-1)$ becomes lower than $E(S=0,S^{z}=0)$ for $\epsilon>J$. When we take a parameter range in which the triplet (disentangled product state) is stabilized, a viewpoint of classical LLG dynamics would be reasonable for describing the dynamics of $\vec{S}$. In this proceeding, we would like to start with such a simple case, and then consider the coupling with bosonic environment. We are interested in a parameter region in which singlet and triplet states are strongly competing with each other, but this is a future work. For comparison, we show that the operator $\vec{m}$ satisfies the following relation
\begin{eqnarray}
\vec{m}\cdot\vec{m}=\frac{3}{8}-\frac{1}{2}\vec{S}_{1}\cdot\vec{S}_{2}=\frac{3}{4}-\frac{1}{4}S(S+1).
\end{eqnarray}
Thus the magnitude of this quantity is also characterized by the total spin $S$. This value for the singlet state with finite amont of entanglement is larger than that for the triplet state. This result also supports that $\vec{m}$ characterizes entanglement between two spins.

For numerical simulation, we take $J=1$, $\omega_{c}=200$, and $\vec{B}_{0}=(h,0,\epsilon)=(0,0,2)$. We introduce the initial quantum state as a product (disentangled) state, and then consider how the two-spin correlation or entanglement is generated by the time evolution:
\begin{eqnarray}
\left|\psi\right>\propto\left(a\left|\uparrow\right>_{1}+(1-a)\left|\downarrow\right>_{1}\right)\otimes\left(b\left|\uparrow\right>_{2}+(1-b)\left|\downarrow\right>_{2}\right)\otimes\left|\varphi\right>,
\end{eqnarray}
where we take $a=0.7$ and $b=0.3$, and $\left|\varphi\right>$ is a bosonic part. For this initial state, the expectation values of $\vec{S}$ and $\vec{m}$ are, respectively, given by
\begin{eqnarray}
\langle\vec{S}\rangle=\left<\psi\right|\vec{S}\left|\psi\right>=\left(0.72, 0, 0\right) \; , \; \langle\vec{m}\rangle=\left<\psi\right|\vec{m}\left|\psi\right>=\left(0, -0.25, 0\right).
\end{eqnarray}
Note that the magnitudes of these vectors depend on the selection of the initial quantum state $\left|\psi\right>$. We do not take a full polarized state $\left|\downarrow\downarrow\right>$ (or the maximally-entangled singlet state) at $t=0$, since $\langle\vec{m}\rangle$ (or $\langle\vec{S}\rangle$) is zero in this case. Our equations of motion, Eqs.~(\ref{LLGsingle}) and (\ref{mdot}), are operator relations, not classical vector equations. Thus, we must take expectation values by the state $\left|\psi\right>$ in order to introduce graphical representation. In this process, $\vec{S}\times\partial\vec{S}/\partial t$ in Eq.~(\ref{LLGsingle}) and $\vec{m}\times\partial\vec{S}/\partial t$ in Eq.~(\ref{mdot}) are respectively decomposed into two independent terms:
\begin{eqnarray}
\langle\vec{S}\times\frac{\partial}{\partial t}\vec{S}\rangle\sim\langle\vec{S}\rangle\times\frac{\partial}{\partial t}\langle\vec{S}\rangle \; , \; 
\langle\vec{m}\times\frac{\partial}{\partial t}\vec{S}\rangle\sim\langle\vec{m}\rangle\times\frac{\partial}{\partial t}\langle\vec{S}\rangle.
\end{eqnarray}
The dynamics of $\langle\vec{S}\rangle$ after this approximation becomes equivalent to the classical LLG equation, except that $\langle\vec{B}\rangle$ still contains information of bosons.  Here we neglect time dependence on $\vec{B}(t)$ that appears as a result of the second term in Eq.~(\ref{field}). In this case, we can solve the equation of motion, and we find that the relaxation time scale $T_{1}$ is proportional to $(1+\pi^{2}\eta^{2}|\langle\vec{S}\rangle|^{2})/2\pi\eta\epsilon|\langle\vec{S}\rangle|$. Unfortunately, quantum effects originated from non-commutativity between $\vec{S}$ and $\partial\vec{S}/\partial t$ and polaronic effects are lost in this approximation, and then $|\langle\vec{S}\rangle|$ is kept. Thus, we suppose that the realistic relaxation time scale may change. In the present approximation, the decoherence time $T_{2}$ becomes infinity due to the stability of the dynamics of $\vec{m}$. The precise estimation of $T_{2}$ is an important future work, but we can say $T_{2}>T_{1}$ even within the present simple analysis.

\begin{figure}[tbh]
\begin{center}
\includegraphics[width=8cm]{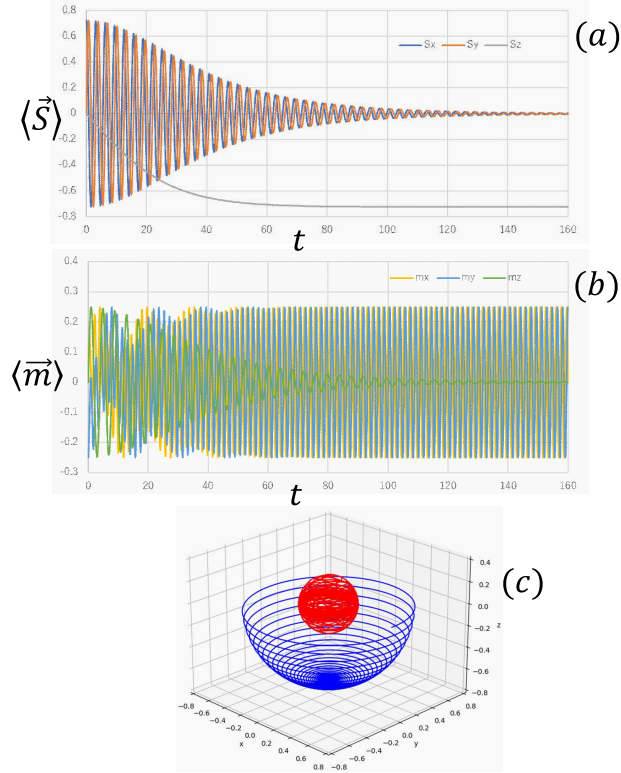}
\end{center}
\caption{Spin dynamics for $\eta=0.008$. (a) $\langle\vec{S}\rangle$, (b) $\left<\vec{m}\right>$, and (c) Graphical representation of (a) and (b). In figure (c), the blue curve represents $\langle\vec{S}\rangle$, and the red curve represents $\left<\vec{m}\right>$.}
\label{f1}
\end{figure}

We demonstrate time evolution of $\langle\vec{S}\rangle$ and $\langle\vec{m}\rangle$ for $\eta=0.008$ ($2\eta\omega_{c}=3.2>J=1$) in Fig.~\ref{f1}. We find that $\langle\vec{S}\rangle$ decays into the direction of $-\langle\vec{B}\rangle$. This feature is consistent with the classical LLG dynamics. As we have already discussed, the length of $\langle\vec{S}\rangle$, $0.72$, is conserved in the present approximation, and $\langle\vec{S}\rangle$ does not become $(0,0,-1)$ even after the long time. We particularly focus on the entanglement dynamics represented by $\langle\vec{m}\rangle$. In contrast to $\langle\vec{S}\rangle$, the coherent oscillation of $\langle\vec{m}\rangle$ is maintained even after $\langle\vec{S}\rangle$ has been relaxed to the stationary point. The coherent oscillation corresponds to continuous spin flip ($\left|\uparrow\downarrow\right>\leftrightarrow\left|\downarrow\uparrow\right>$) between two spins. We find that the phase difference between $\langle m^{x}\rangle$ and $\langle m^{y}\rangle$ is $\pi/2$. As we have already mentioned, the equation of motion for $\vec{m}$ does not contain the damping term if $\partial S/\partial t$ becomes zero. This is the origin of the stable oscillation of $\langle\vec{m}\rangle$. Therefore, the two-spin dynamics is essentially different from total-spin dynamics.

\section{Concluding Remarks}

We derived the spin dynamics in the spin-boson model in order to examine the entanglement control of qubits against dissipation due to the presence of the environment. For the total spin, we found that the result is consistent with the LLG equation for macroscopic spin precession in magnetic materials although the result also contains some quantum effects. Here, the Gilbert constant is proportional to the coefficient of the distribution function of the bosonic degrees of freedom. However, the entanglement dynamics originated from a composite spin shows different behavior. We numerically showed the relaxation dynamics of the total spin and the stability of the entanglement dynamics. In the present approximation, the decoherence time $T_{2}$ is infinity, and more precise treatment based on the higher-order equation of motion is an interesting future work.

H.M. is supported by JPSJ KAKENHI (Nos. 21K03380, 21H04446, 21H03455) from MEXT Japan and CSIS, Tohoku University, Japan. S.M is supported by JST CREST Grant (Nos. JPMJCR19J4, JPMJCR1874, and JPMJCR20C1) and JSPJ KAKENHI (Nos. 17H02927 and 20H01865) from MEXT, Japan.

\end{document}